\documentclass[preprint,showpacs,preprintnumbers,amsmath,amssymb]{revtex4}
\usepackage{graphicx}
\usepackage{dcolumn}
\begin{document}
\preprint{preprint - Low Temperature Physics Group, IIT Bombay}
\title{Enhanced superconducting properties in FeCr$_x$Se}
\author{Anil K. Yadav, Ajay D. Thakur\footnote{ajay@phy.iitb.ac.in}, C. V. Tomy}


\affiliation{Department of Physics, IIT Bombay, Mumbai 400076, India}

\begin{abstract}
We report an enhancement of superconducting transition temperature ($T_{\rm c}$) when Chromium (Cr) is substituted in excess at the Iron (Fe) site (FeCr$_x$Se, $x=$~0.01, 0.02 and 0.03). There is a corresponding increase in the superconducting volume fraction with $T_{\rm c}$ attaining a value of 11.2~K  on 2~$\%$ Cr substitution when compared to a $T_{\rm c}$ of 8.5~K for the conventional tetragonal Fe-excess sample Fe$_{1+x}$Se. The results point to the role of chemical pressure (introduced via ionic size variation at the Fe site upon Cr substitution in excess) on superconducting properties. 
\end{abstract}

\maketitle

\section{Introduction}
There has been a concerted research effort dedicated on Fe-based pnictide  and chalcogenide superconductors since their discovery in early 2008 \cite{kami, rotter, hsu}. The focus of the ongoing investigations has been two-fold : (i) understanding the mechanism of superconductivity in these classes of materials, and (ii) efforts to enhance the transition temperatures. Among these materials, FeSe and its derivatives (commonly known as the `11' system) have drawn immense attention \cite{hsu, takano}. Its Fermi surface is very similar to that reported for the FeAs based superconductors \cite{fermi} comprising of cylindrical electron sections at the zone corners, cylindrical hole surface sections, and small hole sections at the zone center. Furthermore, these surfaces are separated by a 2D nesting vector at ($\pi$, $\pi$), another riminiscent of the FeAs based superconductors. Therefore, akin to their structural and compositional simplicity compared to the pnictide counterparts, they are conceived to be promising model systems to understand the physics of Fe-based superconductors.  Structurally, FeSe comprises of stacked layers of corner sharing FeSe$_4$ tetrahedra similar to FeAs based materials. However, the spacer layers are absent. Analogous to the members of the pnictide family, there is a distortion in the FeSe layers across the structural transition. The stoichiometric members $\alpha$-FeSe and $\alpha$-FeTe (hexagonal NiAs type crystal structure with P6$_3$/mmc space group) are non-superconducting.  A slight excess of Fe is found to stabilize the superconducting tetragonal phase $\beta$-Fe$_{1+x}$Se (with PbO type tetragonal structure and $P4/nmm$ space group) with a $T_{\rm c}\sim$ 8\,K. However, superconductivity
is observed only in a very narrow composition range Fe$_{1.01}$Se - Fe$_{1.04}$Se for the tetragonal phase, reported more than three decades ago \citet{beta1,beta2}. Substitution of Te for Se is found to be increasing the $T_{c}$ in FeSe$_{1-x}$Te$_{x}$ compounds with a maximum $T_{\rm c}$ at $x=0.5$ \citet{fesete1,fesete2}. However, S substitution is found to increase only the superconducting volume fraction \citet{feses1,feses2,feses3} rather than the $T_{\rm c}$. 

The properties of off-stoichiometric $\beta$-Fe$_{1+x}$Se (or, $\beta$-FeSe$_{1-x}$) is very sensitive to the ambient pressure. Mizuguchi {\it et al} \cite{mizu0} reported a $T_{\rm c}$ of 27~K at 1.48 GPa in $\beta$-FeSe. Medvedev {\it et al} \cite{med} found a non-monoticity in the T$_c$ evolution with pressure with a maximum $T_{\rm c}$ of 36 K at 8.9 GPa. After that $T_{\rm c}$ was seen to  decrease with further increase in pressure up to 32 GPa. A similar non-monotonic variation of $T_{\rm c}$ was observed in high pressure studies by Margadonna {\it et al} \cite{mar} where T$_c$ was seen to peak at 37 K at 7 GPa. In contrast, Garbarino {\it et al} \cite{gar} showed a monotonic increase in $T_{\rm c}$ with pressure. In their studies, an orthorhombic high pressure phase was seen to develop at 12 GPa with a T$_c$ of 34 K at 22 GPa. Although the exact nature of the superconducting phase is not very clear, the sensitive dependence of $T_{\rm c}$ on ambient pressure suggests a possibility of increasing $T_{\rm c}$ via the chemical pressure route. Different substitutions at the Fe site have been tried in the past by several groups \cite{tm1, tm2, tm3} however, there has been no reports suggesting an enhancement of $T_{\rm c}$. The superconducting volume fraction was seen to enhance on Ni substitution at Fe site \cite{ni1} but there was no concomitant increase in $T_{\rm c}$. We report here the effect excess Cr at the Fe site on superconducting properties. The results suggest an enhancement of T$_c$ with accompanied enhancement in superconducting volume fraction.
\section{Experimental}
Polycrystalline samples of FeCr$_x$Se ($x = 0.01, 0.02$) were prepared using conventional solid state reaction technique. Reactants were homogenized in an agate mortar; it was packed and sealed in a quartz tube and preheated for 24 hrs at 500$^o$C, reground, sealed and heated at 1050$^o$C for 24 hours. A batch of samples were post-annealed at 360$^o$C for 3h hours. Using powder X-ray diffraction (XRD), the samples were characterized for phase purity. Magnetic and transport studies on the samples were performed using Quantum Design Inc. PPMS. During transport measurements, the dc current was maintained so as to achieve 100 $\mu$V at room temperature across the voltage leads in the four probe configuration. During magnetic measurements using the PPMS vibrating sample magnetometer (VSM), the amplitude of vibration was set as 2.0 mm at 40 Hz.

\section{Results and Discussion}

The powder XRD patterns for FeCr$_x$Se ($x =$~0.01, 0.02 and 0.03) samples are shown in Fig. 1. These patterns were indexed according to the Tetragonal $\beta$-FeSe with P4/nmm space group. Certain peaks corresponding to the impurity phase of hexagonal $\alpha$-FeSe (NiAs structure) with P6$_3$/mmc space group (labeled as $\alpha$) were also detected. For comparison purposes, the XRD pattern corresponding to the Fe-excess sample, Fe$_{1.01}$Se is also plotted in Fig. 1.  Contributions from other impurity phases like Fe$_{3}$O$_{4}$ were negligible. Following observations are noteworthy: (i) with increase in Cr concentration, there is a notable shifting of the peaks to smaller 2$\theta$ values indicating a gradual expansion of the unit cell volume (see inset (a) and (b) in Fig. 1), (ii) prominent peaks corresponding to the hexagonal $\alpha$ phase are seen to evolve with increase in Cr concentration.

Figure~ 2 shows the magnetization measurement as a function of temperature ($M$-$T$ plots) for the FeCr$_{x}$Se ($x=$~0.01, 0.02 and 0.03) samples. The data is taken in both the zero field-cooled (ZFC) as well as the field-cooled (FC) mode. The onset of the superconducting transition is marked by arrows and its values are 10.4\,K, 11.2\,K
and 11.2\,K, respectively, for $x=$~0.01, 0.02 and 0.03 samples. As can be inferred from Fig.~2, the superconducting volume fraction is maximum for the $x=$~0.02 composition. To the best of our knowledge, the $T_{{\rm c}}$ for our samples with Cr substitution exceeds the maximum reported $T_{{\rm c}}$ in the FeSe system with substitutions
at Fe site (including the Fe-excess samples) \citet{hsu,takano,tm1,tm2,tm3,ni1}. In the past, Wu \textit{et al} \citet{tm1} have reported results on Cr substitution at the Fe site, but did not see any dramatic $T_{{\rm c}}$ enhancement. This may be due to the fact that they have added an excess of 10\% Cr ($x=0.1$) which is far greater than the value $x=0.03$
in our case. It is worthwhile to recall the work by Ge \textit{et al} \citet{ge}, where they observed a $T_{{\rm c}}$ of 10.9\,K at ambient pressure in Fe$_{1.03}$Se system arising because of the internal crystallographic strain. However, in their case, post annealing treatment led to a substantial decrease in $T_{{\rm c}}$. In our case, the $T_{{\rm c}}$ remains $\sim$11\,K with negligible changes to superconducting volume fractions despite a similar post annealing protocol (data not shown) pointing to the role of Cr substitution
in enhancing the internal chemical pressure and thereby leading to higher $T_{{\rm c}}$. Future work on similar lines could provide routes to achieve higher $T_{{\rm c}}$ at ambient pressures in other Fe-based superconductors.

In Fig.~3, we show a comparison of the $T_{{\rm c}}$'s obtained via magnetization measurements and dc transport studies in the $x=0.02$ sample. As can be seen, the temperature for the onset of diamagnetism ($T_{{\rm c}}^{M}$) coincide with the temperature at which the resistance falls to 10 $\%$ of the resistance ($T_{{\rm c}}^{R,~offset}$) in the normal state across the normal state to superconducting state transition. This temperature is 11.2\,K. Even though the hexagonal impurity phase is present in these compounds (as inferred from XRD in Fig.~1), the insensitivity of $T_{c}$ to the presence of this additional phase reaffirms the recent observations made by Tsurkan \textit{et al} \citet{tsurkan} in FeSe$_{0.5}$Te$_{0.5}$. Also marked in Fig.~3 are the $T_{{\rm c}}^{R,~onset}$, and $T_{{\rm c}}^{R,~mid}$ where the resistance reaches 90\% and 50\% of the normal state value, respectively. The inset in Fig.~3 shows the resistance as a function of temperature in the temperature range 2\,K to 300\,K. The residual resistivity ratio (RRR) given by $\frac{R_{300~K}}{R_{12~K}}$ for the $x=0.02$ sample turns out to be $\approx$\,4. In Fig.~4 we show the isothermal magnetization ($M$-$H$) measurement in $x=0.02$
sample at 2\,K (far below $T_{{\rm c}}$) and 15\,K (above $T_{{\rm c}}$). Note the diamagnetic response at low fields in the 2\,K data. This, along with the shape of the $M$-$H$ loop at 2\,K, is characteristic of a typical type-II superconductor. The open loop behaviour at 15\,K may be due to the small amount of impurity phases like Fe$_{3}$O$_{4}$ or Fe, which are not detected in XRD, but is generally present in the FeSe samples.

Specific heat measurements is conventionally used as an equilibrium measurement for demonstrating the existence of bulk superconductivity in samples. In Fig.~5, we show the specific heat data for the $x=0.02$ sample. Here, $C/T$ is plotted as a function of $T^{2}$ in the main panel. The characteristic jump in the value of $C$ is not observed distinctly. Total specific heat primarily comprises lattice, electronic as well as magnetic contributions. In addition, it may also have a contribution from the frozen-in disorder in the sample. In order to observe a clear signature of the jump in specific heat due superconductivity, the electronic (normal) and lattice specific heats must be subtracted from the total heat capacity data. One of the common ways to remove the lattice and electronic contributions is to subtract the specific heat of a non-superconducting iso-structural system. We have observed that the addition of excess Co at the Fe site (FeCo$_{x}$Se) suppresses superconductivity \cite{aky}. Hence we have subtracted the specific heat of the compound FeCo$_{0.01}$Se, for removing the approximate lattice and electronic contributions from FeCr$_{0.02}$Se. In addition, this process may also eliminate the approximate magnetic as well as the frozen-in disorder contributions to the total specific heat. In the inset of Fig.~5, we show such a difference plot for $\Delta C/T$ versus $T^{2}$. Note that a characteristic jump is visible across the superconducting transition. As the method used is only an approximate estimation, we have not attempted to estimate the superconducting energy gap ($\Delta_{SC}$). However, our method does indicate the existence of bulk superconductivity in FeCr$_{0.02}$Se.

Finally, we make an estimation of the upper critical field, $H_{\rm c2}$ for polycrystalline FeCr$_{0.02}$Se. In Fig. 6, we show a plot of $T_{\rm c}^{R, onset}$, $T_{\rm c}^{R, mid}$ and $T_{\rm c}^{R, offset}$ obtained from magneto-transport measurements. Within, the formulation suggested by Werthamer {\it et al} \cite{whh} (well known as the WHH theory), $H_{\rm c2}$ is given by,  $\mu_{\rm 0}H_{c2} \left( 0 \right) = -0.693~\mu_{\rm 0}\left( \frac{dH_{\rm c2}}{dT}\right)_{T_{\rm c}} T_{\rm c}$. From Fig. 6, the slopes $\left( \frac{dH_{\rm c2}}{dT}\right)_{T_{\rm c}}$ are obtained and we make an estimation of $H_{\rm c2}$. The $H_{\rm c2}(0)$ values obtained from $T_{\rm c}^{R, onset}$, $T_{\rm c}^{R, mid}$ and $T_{\rm c}^{R, offset}$ data are 37.47~T, 23.66~T and 19.03~T, respectively. The Ginzburg-Landau expression for the coherence length ($\xi$) is given by, $\xi (0) = \left( \frac{\Phi_{\rm 0}}{2\pi \mu_{\rm 0} H_{\rm c2}(0)} \right)^{\frac{1}{2}}$, where, $\Phi_{\rm 0}$ is the quantum of flux with a magnitude 2.07$\times$10$^{-7}$ Gauss-cm$^2$. This leads to $\xi (0)$ of 29.6~$\AA$, 37.3~$\AA$ and 41.6~$\AA$ for the data corresponding to onset, mid and offset positions, respectively. Here, it should be kept in mind that the $H_{c2}$ and $\xi$ values obtained here corresponds to the polycrystalline samples. For studying the role of crystalline anisotropy, single crystals needs to be prepared for the Cr excess samples. Further work is in progress in this direction.

\section{Summary}
To summarize, we observe an enhancement in T$_c$ in samples with Cr excess at the Fe site in parent FeSe system at ambient pressure. The results points to the importance of chemical pressure achieved via suitable site substitution as a route to enhancement of superconducting properties in Fe-based superconducting materials. We believe that our results would stimulate further research on the `11' system.

\section*{Acknowledgements}
CVT would like to acknowledge the Department of Science and Technology for partial support through the project IR/S2/PU-10/2006. ADT would like to acknowledge the Institute Post Doctoral Fellowship at the Indian Institute of Technology, Bombay for partial support.

\newpage

\begin{figure}
\includegraphics[scale=0.5,angle=0]{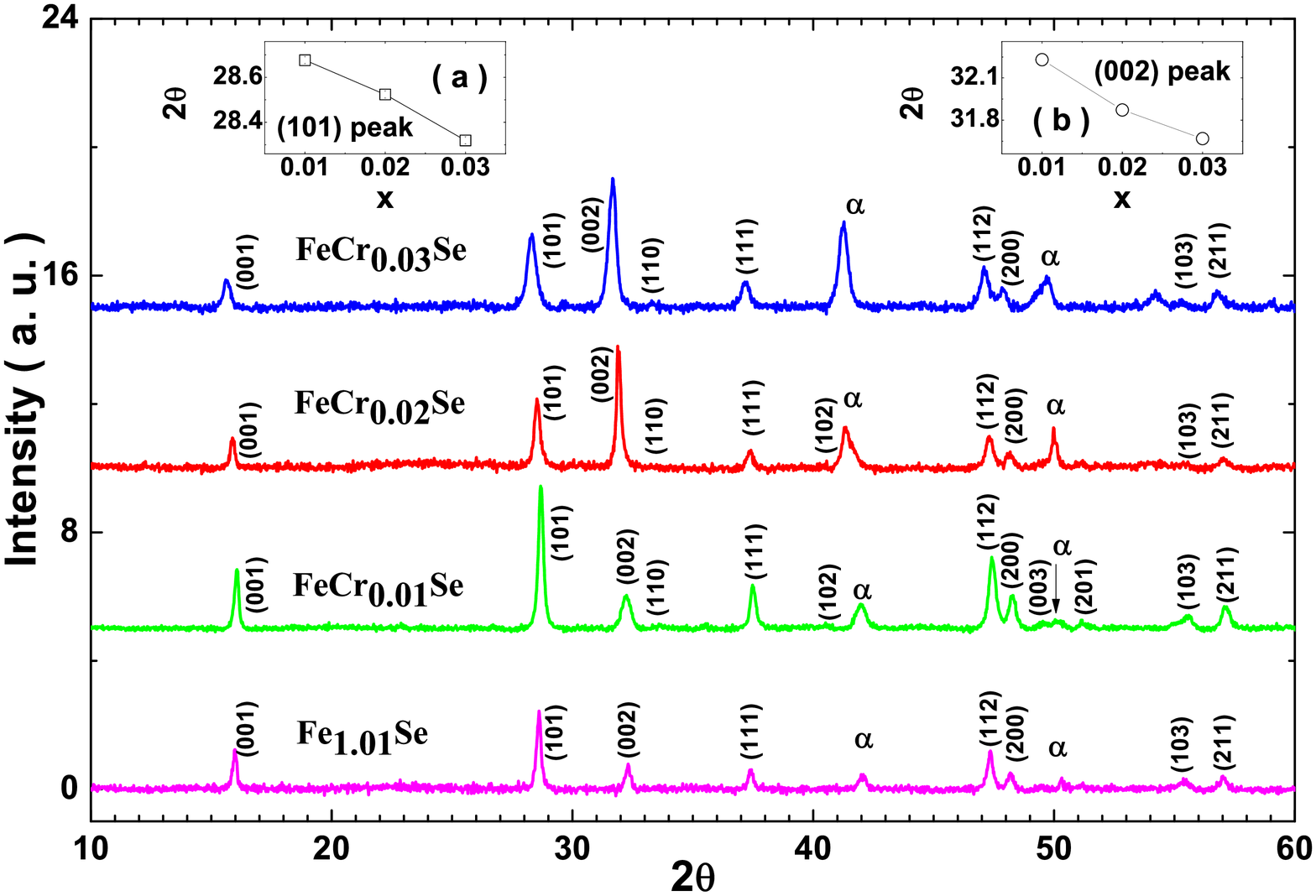}
\caption{(Color online) The powder XRD patterns for FeCr$_x$Se ($x =$~0.01, 0.02 and 0.03) samples obtained using Cu-K$_\alpha$ x-rays. Also shown is the XRD spectra for Fe$_{1.01}$Se. The peaks corresponding to the hexagonal phase with P6$_3$/mmc space group are marked by $\alpha$. Insets (a) and (b) shows the evolution of the $2\theta$ values for the (101) and (002) peaks, respectively as a function of x.}
\end{figure}

\begin{figure}
\includegraphics[scale=0.5,angle=0]{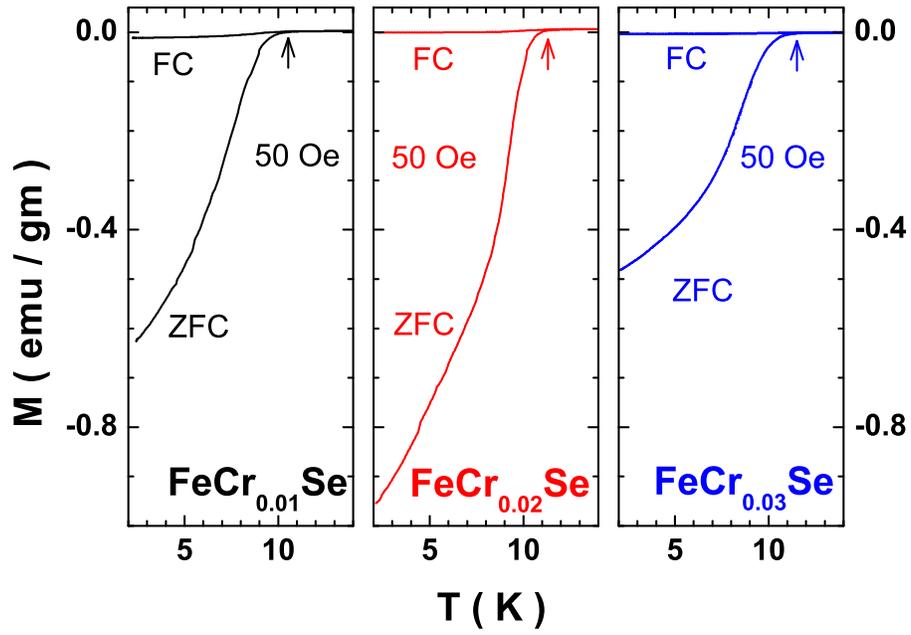}
\caption{(Color online) Magnetization measurement as a function of temperature (M-T plots) for the FeCr$_x$Se ($x =$~0.01, 0.02 and 0.03) samples at an applied dc magnetic field of 50 Oe. The zero field cooled (ZFC) and the field cooled (FC) runs are labeled..}
\end{figure}

\begin{figure}
\includegraphics[scale=0.6,angle=0]{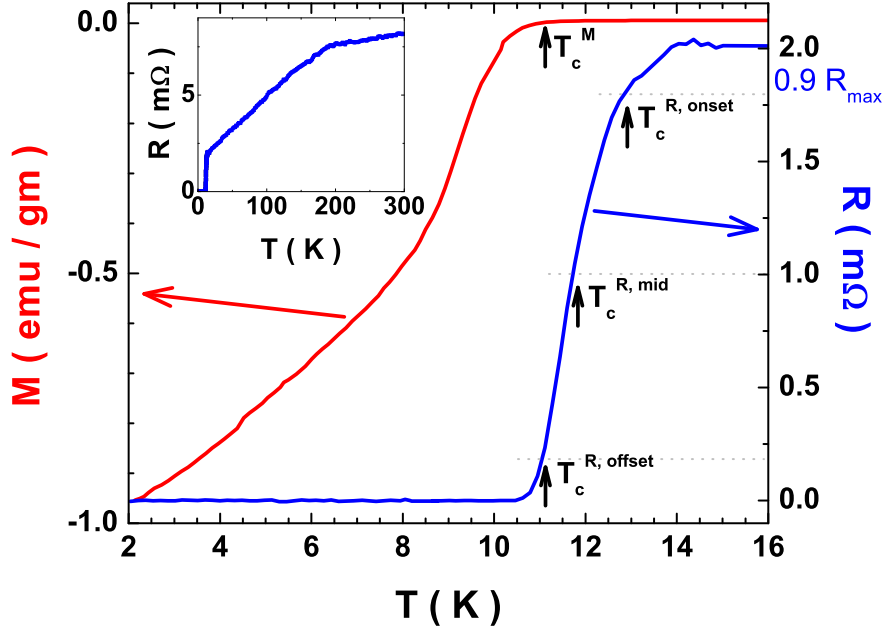}
\caption{(Color online) Comparison of magnetization (M-T) and transport (R-T) data for obtaining the $T_{\rm c}$ in the case of FeCr$_{0.02}$Se. The locations of $T_{\rm c}^M$, $T_{\rm c}^{R,~onset}$, $T_{\rm c}^{R,~mid}$ and $T_{\rm c}^{R,~offset}$ are marked by arrows. The inset shows the R-T data from 300~K to 2~K}
\end{figure}

\begin{figure}
\includegraphics[scale=0.6,angle=0]{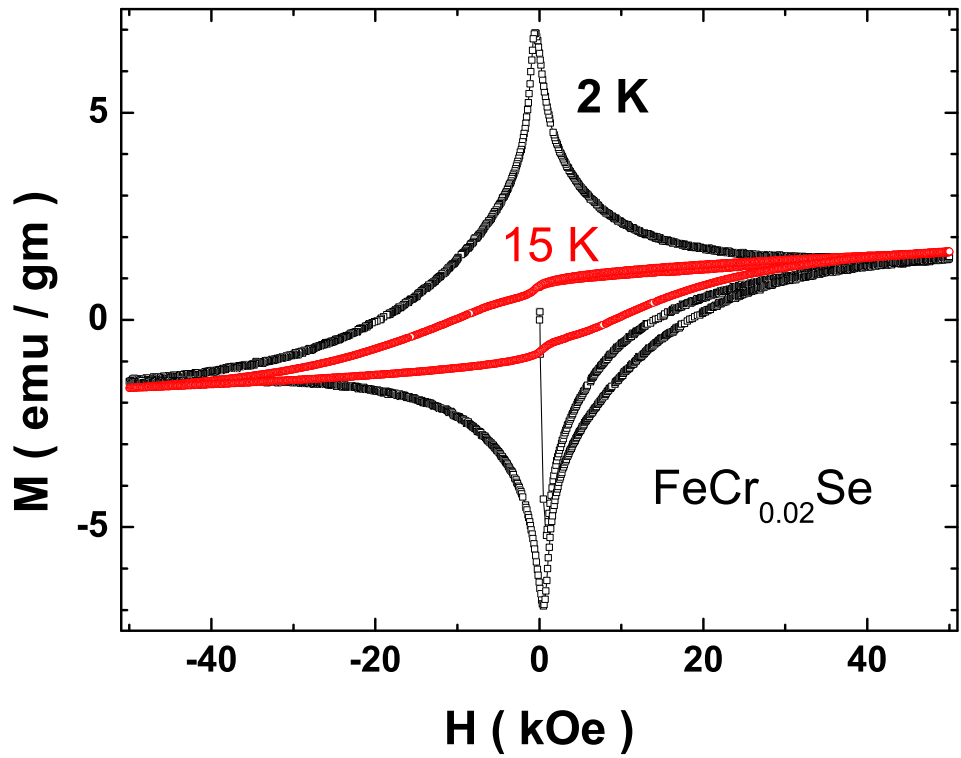}
\caption{(Color online) Isothermal magnetization hysteresis (M-H) loop for FeCr$_{0.02}$Se obtained at 2~K (open black squares)  and 15~K (open red squares).}
\end{figure}

\begin{figure}
\includegraphics[scale=0.6,angle=0]{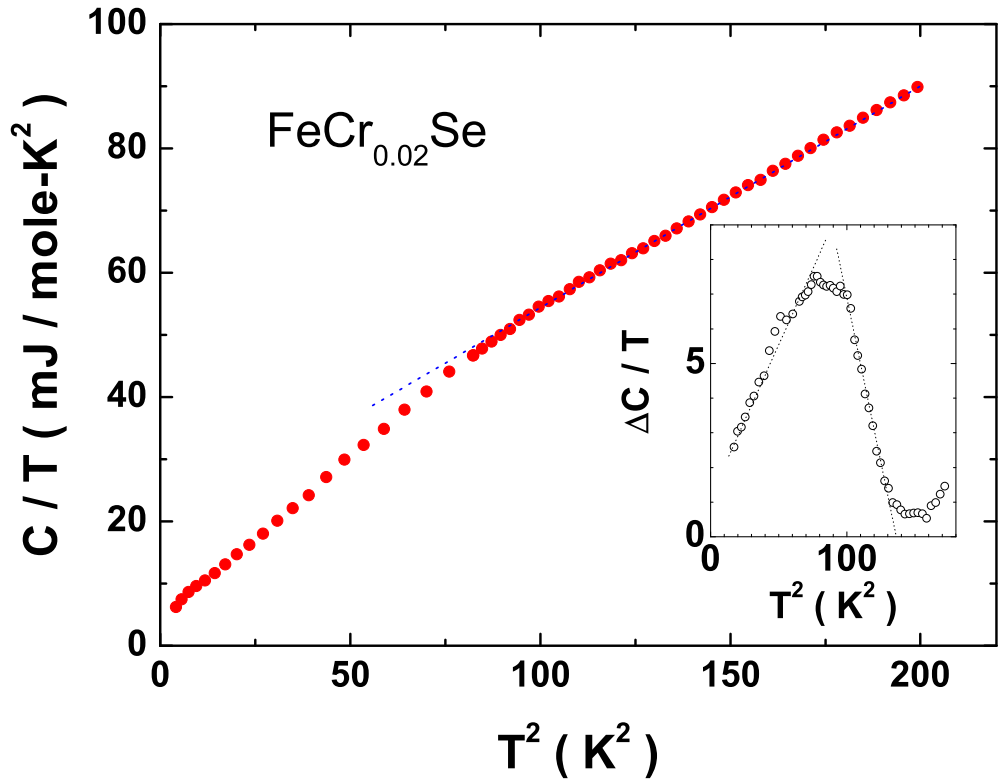}
\caption{(Color online) $C/T$ versus $T^2$ for FeCr$_{0.02}$Se. The inset shows a plot of $\Delta C/T$ versus $T^2$ (see text for details).}
\end{figure}

\begin{figure}
\includegraphics[scale=0.6,angle=0]{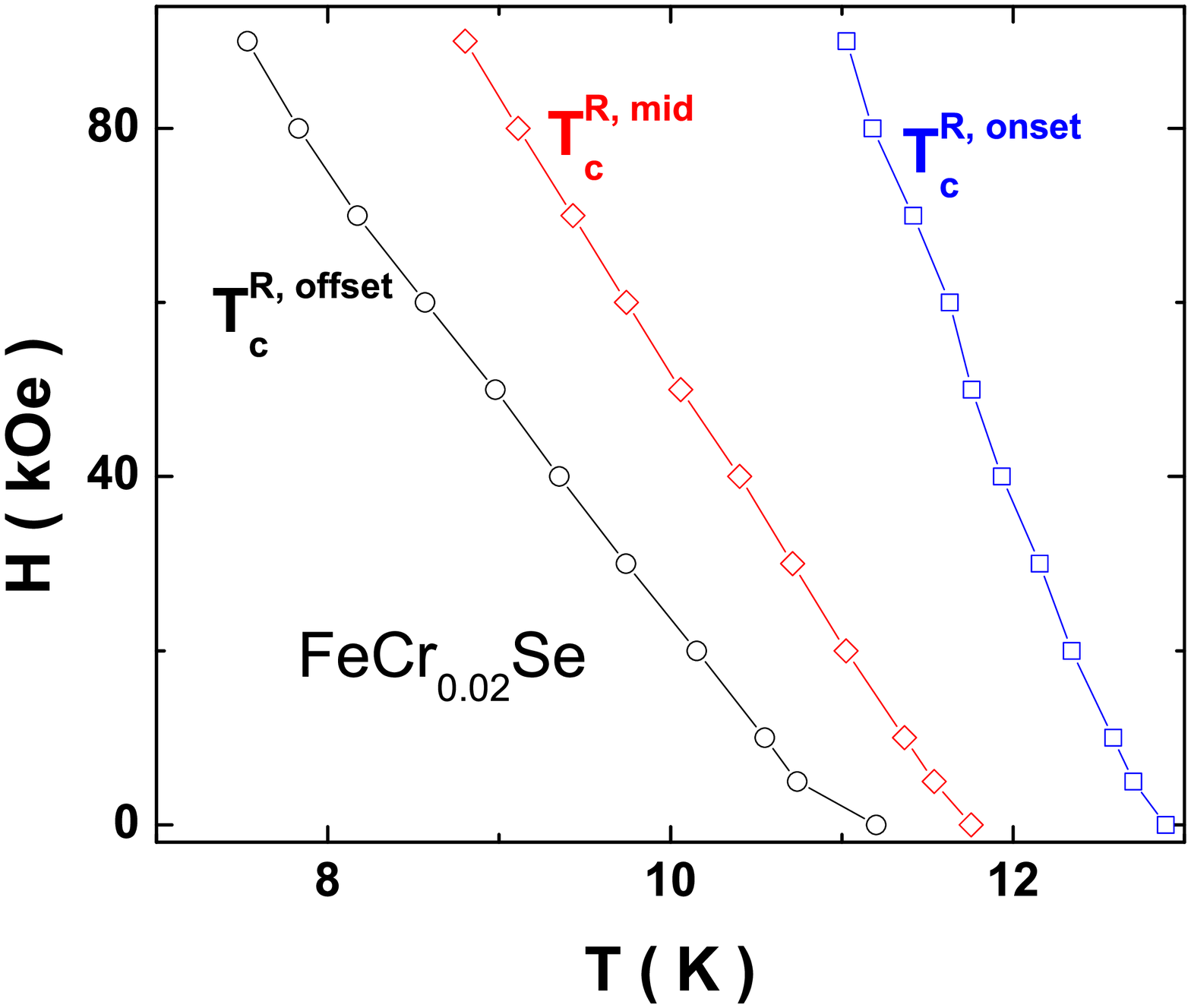}
\caption{(Color online) Phase diagram for the $x=0.02$ sample obtained via transport studies. $T_{\rm c}^{R, onset}$, $T_{\rm c}^{R, mid}$ and $T_{\rm c}^{R, offset}$ are plotted as a function of field.}
\end{figure}


\begin{thebibliography}{00}

\bibitem{kami}
Y.Kamihara, H.Hiramtsu, M.Hirano, R.Kawamura, H.Yanagi, T. Kamiya, and H.Hosono,  J. Am. Chem. Soc. 130 (2008) 3296.

\bibitem{rotter}
M. Rotter,  M. Tegel, D. Johrendt, Phys Rev. Lett. {\bf 101} (2008), p. 026403.

\bibitem{hsu}
F.C. Hsu, J.Y. Luo, K.W. Yeh, T.K. Chen, T.W. Huang, P.M. Wu, Y.C. Lee, Y.L. Huang, Y.Y. Chu, D.C. Yan, M.K. Wu, Proc. Natl. Acad. Sci. USA 105 (2008) 14262.

\bibitem{takano}
Y. Mizuguchi and Y. Yakano, Journal of the Physical Society of Japan, 79 (2010) 102001.

\bibitem{fermi}
A. Subedi, L. Zhang, D. J. Singh, and M. H. Du, Phys. Rev. B 78 (2008) 134514. 

\bibitem{beta1}
W. Schuster, H. Mikler, and K. L. Komarek, Monatsch. Chem. 110 (1979) 1153.

\bibitem{beta2}
H. Okamoto, J. Phase Equilib. 12 (1991) 383.

\bibitem{fesete1}
M. H. Fang, H. M. Pham, B. Qian, T. J. Liu, E. K. Vehstedt, Y. Liu, L. Spinu and Z. Q. Mao, Phys. Rev. B 78 (2008) 224503.

\bibitem{fesete2}
K. W. Yeh, T. W. Huang, Y. Huang, T. K. Chen, F. C. Hsu, P. M. Wu, Y. C. Lee, Y. Y. Chu, C. L. Chen, J. Y. Luo, D. C. Yan and M. K. Wu, Europhys. Lett. 84 (2008) 37002.

\bibitem{feses1}
Y. Mizuguchi, F. Tomioka, S. Tsuda, T. Yamaguchi and Y. Takano, J. Phys. Soc. Japan 78 (2009) 074712.

\bibitem{feses2}
R. Hu, E. S. Bozin, J. B. Warren and C. Petrovic, Phys. Rev. B 80 (2009) 214514.

\bibitem{feses3}
Y. Mizuguchi, F. Tomioka, S. Tsuda, T. Yamaguchi and Y. Takano, Appl. Phys. Lett. 94 (2009) 012503.

\bibitem{mizu0}
Y. Mizuguchi, F. Tomioka, S. Tsuda, T. Yamaguchi and Y. Takano, Appl. Phys. Lett. 91 (2008) 152505.

\bibitem{med}
S. Medvedev, T.M. McQueen, I. Trojan, T. Palasyuk, M.I. Eremets, R.J. Cava, S. Naghavi, F. Casper, V. Ksenofontov, G. Wortmann, C. Felser, Nature Materials 8, 630 (2009).

\bibitem{mar}
S. Margadonna, Y. Takabayashi, M. T. McDonald, K. Kasperkiewicz, Y. Mizuguchi, Y. Takano, A. N. Fitch, E. Suard and K. Prassides, Chem. Commun., 5607 (2008).

\bibitem{gar}
G. Garbarino, A. Sow, P. Lejay, A. Sulpice, P. Toulemonde, M. Mezouar and M. N$\acute{u}\tilde{n}$ez-Regueiro, Euro. Phys. Lett. 86, 27001 (2009).

\bibitem{tm1}
M. K. Wu, F.C. Hsu, K.W. Yeh, T.W. Huang, J.Y. Luo, M.J. Wang, H.H. Chang, T.K. Chen, S.M. Rao, B.H. Mok, C.L. Chen, Y.L. Huang, C.T. Ke, P.M. Wu, A.M. Chang, C.T. Wu, T.P. Perng, Physica C 469, 340 (2009).

\bibitem{tm2}
Y. Mizuguchi, F. Tomioka, S. Tsuda, T. Yamaguchi and Y. Takano,  J. Phys. Soc. Japan 78, 074712 (2009).

\bibitem{tm3}
D. J. Gawryluka, J. Fink-Finowickia, A. Wisniewskia, R. Puzniaka, V. Domukhovskia, R. Diduszkoa,b, M. Kozlowskia, and M. Berkowskia, arXiv : cond-mat/1010.4217.

\bibitem{ni1}
A. G$\ddot{u}$nther, J. Deisenhofer, Ch. Kant, H.-A. Krug von Nidda, V. Tsurkan, and A. Loidl, arXiv : cond-mat/1010.5597.

\bibitem{ge}
J. Ge, S. Cao, S. Yuan, B. Kang, and J. Zhang, J. Appl. Phys. 108 (2010) 053903.

\bibitem{tsurkan}
V. Tsurkan, J. Deisenhofer, A. G$\ddot{u}$nther, Ch. Kant, H.-A. Krug von Nidda, F. Schrettle, and A. Loidl, arXiv : cond-mat/1006.4453v2.

\bibitem{aky}
Anil K. Yadav {\it et al} (to be published).

\bibitem{whh}
N. R. Werthamer, E. Helfand, P. C. Hohenberg, Phys. Rev. 147 (1966) 295.

\end{thebibliography}
\end{document}